\documentclass[aps,twocolumn,superscriptaddress,floatfix,longbibliography]{revtex4-2}
\usepackage{graphicx,amssymb,amsmath,mathrsfs,hyperref,bm,dsfont,xcolor}

\usepackage{subfigure}
\newif\ifhyper
\hypertrue
\ifhyper
\hypersetup{
   citecolor = {red},
   colorlinks = {true}, 
   linkcolor = {blue},
   urlcolor = {blue} 
}
\fi

\def\be{\begin{equation}}
\def\ee{\end{equation}}
\def\bea{\begin{eqnarray}}
\def\eea{\end{eqnarray}}

\newcommand{\ket}[1]{|#1\rangle}

\begin{document}

\title{Quantum Phase Diagram of the Bilayer Kitaev-Heisenberg Model}

\author{Elahe Samimi}
\affiliation{Department of Physics, Institute for Advanced Studies in Basic Sciences (IASBS), Zanjan 45137-66731, Iran}

\author{Antonia Duft}
\affiliation{Department Physik, Staudtstra{\ss}e 7, Universit\"at Erlangen-N\"urnberg, D-91058 Erlangen, Germany}

\author{Patrick Adelhardt}
\affiliation{Department Physik, Staudtstra{\ss}e 7, Universit\"at Erlangen-N\"urnberg, D-91058 Erlangen, Germany}
\affiliation{Joint Quantum Institute and Joint Center for Quantum Information and Computer Science, University of Maryland Department of Physics and National Institute of Standards and Technology, College Park, Maryland 20742, USA}

\author{Kai Phillip Schmidt}
\affiliation{Department Physik, Staudtstra{\ss}e 7, Universit\"at Erlangen-N\"urnberg, D-91058 Erlangen, Germany}

\author{Saeed S. Jahromi}
\email{saeed.jahromi@iasbs.ac.ir}
\affiliation{Department of Physics, Institute for Advanced Studies in Basic Sciences (IASBS), Zanjan 45137-66731, Iran}

\begin{abstract}

We study the ground-state phase diagram of the spin-$1/2$ Kitaev-Heisenberg model on the bilayer honeycomb lattice with large-scale tensor network calculations based on the infinite projected entangled pair state technique as well as high-order series expansions. We find that beyond various magnetically ordered phases, including ferromagnetic, zigzag, antiferromagnetic (AFM) and stripy states, two extended quantum spin liquid phases arise in the proximity of the Kitaev limit. While these ordered phases also appear in the monolayer Kitaev-Heisenberg model, our results further show that a valence bond solid state emerges in a relatively narrow range of parameter space between the AFM and stripy phases, which can be adiabatically connected to isolated Heisenberg dimers. Our results highlight the importance of considering interlayer interactions on the emergence of novel quantum phases in the bilayer Kitaev materials.

\end{abstract}

\maketitle

\section{Introduction.}

Magnetic frustration is the root to many exotic phenomena in condensed matter physics. Novel quantum phases, such as quantum spin liquids (QSL) \cite{Savary2017,Balents2010}, resonating valence bond (RVB) states \cite{Anderson1973153, Moessner2001}, and valence bond crystals (VBC)  \cite{Iqbal2019,jahromi_spin-frac12_2018, mila_frustrated_magnetism} emerge as a result of quantum fluctuations of frustrated spins in the magnetic systems. QSLs are spin-disordered phases of matter that exhibit emergent properties, such as long-range entanglement, fractionalized excitations, and topological order. These distinctive properties originate from the underlying topological characteristics of the ground state of many-body systems, rendering them promising candidates for the realization of topological quantum computation \cite{Savary2017,Balents2010}. The RVB states, that were originally introduced by Philip Anderson in the context of high-temperature superconductivity \cite{anderson1987resonating}, are another family of QSLs composed of an equal-weight superposition of valence bond coverings (spin singlets) on a given lattice with vanishing local magnetization and no broken symmetry \cite{mila_frustrated_magnetism,fazekas1974ground,bednorz1986possible}.


One of the most widely recognized models that hosts both gapped and gapless QSLs is the spin-$1/2$ Kitaev model on the honeycomb lattice. The model is composed of anisotropic nearest-neighbor bond-dependent Ising-like interactions and is exactly solvable in the Majorana fermion representation. In specific ranges of Kitaev interactions, where the system’s ground state is gapless, emergent excitations in the form of gapless Majorana fermions appear. Conversely, the gapped phase exhibits a $\mathbb{Z}_2$ topological order characterized by Abelian anyonic statistics \cite{kitaev2006anyons}. However, a more realistic model typically incorporates the symmetric Heisenberg interactions \cite{trebst2022kitaev}. Particularly,  the physics of Iridium-based materials such as Na$_{2}$IrO$_{3}$ and Li$_{2}$IrO$_{3}$ with a honeycomb lattice structure can be explained by the Kitaev-Heisenberg Hamiltonian \cite{chaloupka2010kitaev,chaloupka2013zigzag,price2012critical}. It has been shown that the competition between Kitaev and Heisenberg interactions leads to a rich phase diagram including distinct magnetically ordered phases and two QSL states \cite{chaloupka2010kitaev,chaloupka2013zigzag,price2012critical,iregui2014probing}. On the other hand, the layered structure of some honeycomb lattice compounds may play a significant role in determining their low-energy states \cite{singh2010antiferromagnetic}. While 2D lattice models can effectively capture the underlying physics of frustrated quantum magnetism, the absence of interlayer coupling can limit their potential to fully represent the emergent properties that arise in real-world materials \cite{sengupta2003specific}. In layered structures, the interlayer interactions can introduce additional degrees of freedom leading to novel quantum phases. In this regard, it has been shown that the phase diagram of the Kitaev model in the presence of interlayer Heisenberg coupling consists of a QSL state and a dimer phase \cite{seifert2018bilayer}. When the interlayer coupling is replaced by a special form of four-body interaction, the phase transition occurs between a QSL phase and a short-range RVB state \cite{hwang2023topological}. However, the computational cost of numerical simulations in bilayer lattice models can lead to challenges when simulating more realistic systems \cite{seifert2018bilayer,hwang2023topological,wagner2021two,tomishige2018interlayer}. It is therefore of immediate interest to study the ground state properties of the spin-$1/2$ Kitaev-Heisenberg model with the Heisenberg interlayer interaction in a bilayer honeycomb lattice, in contrast to the monolayer system.   

In this regard, we study a bilayer honeycomb lattice model where spin-$1/2$ particles are located at the vertices of the honeycomb structure. Within each layer, spins interact through the Kitaev-Heisenberg (KH) Hamiltonian, while the honeycomb layers are coupled via Heisenberg interactions. We use the state-of-the-art tensor network method \cite{Orus2014,Orus2014a,Verstraete2008} to simulate the ground state of the model in the thermodynamic limit. The method is based on the iPEPS formalism, a variational tensor network ansatz for ground states in the thermodynamic limit \cite{Orus2009,Phien2015,jahromi_infinite_2018}. In addition, we perform complementary high-order series expansions about the limit of isolated rung dimers where the system realizes a featureless valence bond solid (VBS) of rung singlets. In particular, the gap closing of one-triplon energies signals the breakdown to magnetically ordered phases present in the bilayer Kitaev-Heisenberg model. Our findings reveal that similar to the monolayer model, the competition between Kitaev and Heisenberg interactions gives rise to various magnetically ordered phases, including zigzag, stripy, ferromagnetic (FM), and antiferromagnetic (AFM) phases, as well as two extended QSL states close to the Kitaev limit. Furthermore, our results unveil the presence of a VBS phase sandwiched between the AFM and stripy ordered phases, resulting solely from layering the model with Heisenberg interactions. The VBS is shown to be adiabatically connected to isolated dimers so that it is a valence bond crystal (VBC) which breaks no translational symmetry.

The paper is organized as follows: In Section~\ref{sec:model}, we introduce the bilayer Kitaev-Heisenberg model and review some of its relevant properties. Section~\ref{sec:method} provides information on the tensor network algorithm, including the details of the implementation of iPEPS ansatz for layered models, as well as the high-order series expansion. In Section~\ref{sec:phase_diagram}, we present our results and characterize the full phase diagram of the model. Finally, Section~\ref{sec:conclude} is devoted to conclusion and discussion.

\begin{figure}
\centerline{\includegraphics[width=\columnwidth]{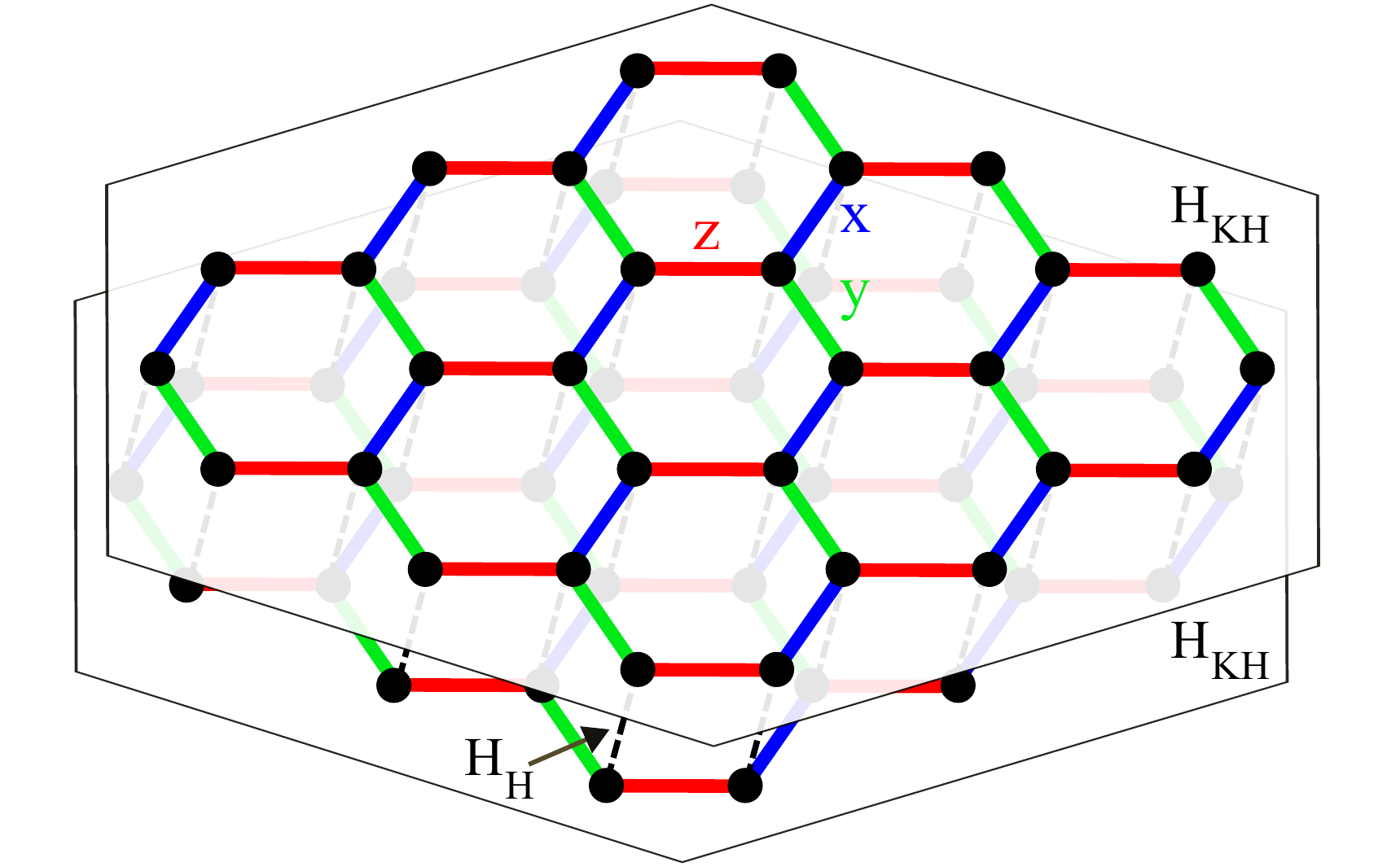}}
\caption{(Color online) Bilayer model with a honeycomb lattice structure. Spin-$1/2$ particles reside on vertices. Bond-directional interactions, denoted by x, y, and z, are distinguished through color-coding in blue, green, and red, respectively. Interlayer interactions, governed by the Heisenberg Hamiltonian $H_H$, are represented by dashed lines. Within each layer, degrees of freedom interact via the Kitaev-Heisenberg Hamiltonian denoted by $H_{KH}$.}
\label{Fig:lattice}
\end{figure}

\section{Model} 
\label{sec:model} 
We study the spin-$1/2$ Kitaev-Heisenberg model on the bilayer honeycomb lattice where honeycomb layers are coupled with Heisenberg interactions. The Hamiltonian of each honeycomb layer is given by
\be
\label{eq:H_AFH}
H_{KH} = \sum_{{\langle i j \rangle}_{\alpha}}K^{\alpha}{S}_{i}^{\alpha}{S}_{j}^{\alpha} +\sum_{\langle i j \rangle} J_{i,j} \, \mathbf{S}_i \cdot \mathbf{S}_j, 
\ee
where the sum runs over nearest-neighbor spins ${\langle i j \rangle}$ of the lattice, $\alpha = x,y,z$, and $S_{i}^{\alpha}$ denotes the spin-$1/2$ operator associated with the $\alpha$-component. $ K^{x}, K^{y}$ and $K^{z}$ are exchange coupling interactions corresponding to the $x, y$ and $z$ links, respectively (see Fig.~\ref{Fig:lattice}). The honeycomb layers are coupled through the Heisenberg Hamiltonian defined as
\be
\label{eq:H_I}
H_{H} = \sum_{i} J_{iu,id}\,\mathbf{S}_{iu} \cdot \mathbf{S}_{id}, 
\ee
where $\mathbf{S}_{iu(id)}$ denotes the spin-$1/2$ operator at site-$i$ of the up and down layers, respectively and $J_{iu,id}$ refers to the strength of the on-site nearest neighbor interlayer Heisenberg interaction between two layers. For the limiting case $J_{i,j}=K^{\alpha}=0$, one obtains isolated rung dimers of nearest-neighbor spins on the two different layers which serves as the unperturbed limit for the high-order series expansions.
Here, we consider the isotropic case for both the Kitaev and Heisenberg interactions with $K^{x}=K^{y}=K^{z}=K$ and $J_{iu,id}=J_{i,j}=J$. 
We further parameterize the couplings as $K=\cos{\theta}$ and $J=\sin{\theta}$ and let $\theta$ vary from $0$ to $2\pi$ and extract the full phase diagram of the model for different ranges of $\theta$.

\begin{figure}
\centerline{\includegraphics[width=\columnwidth]{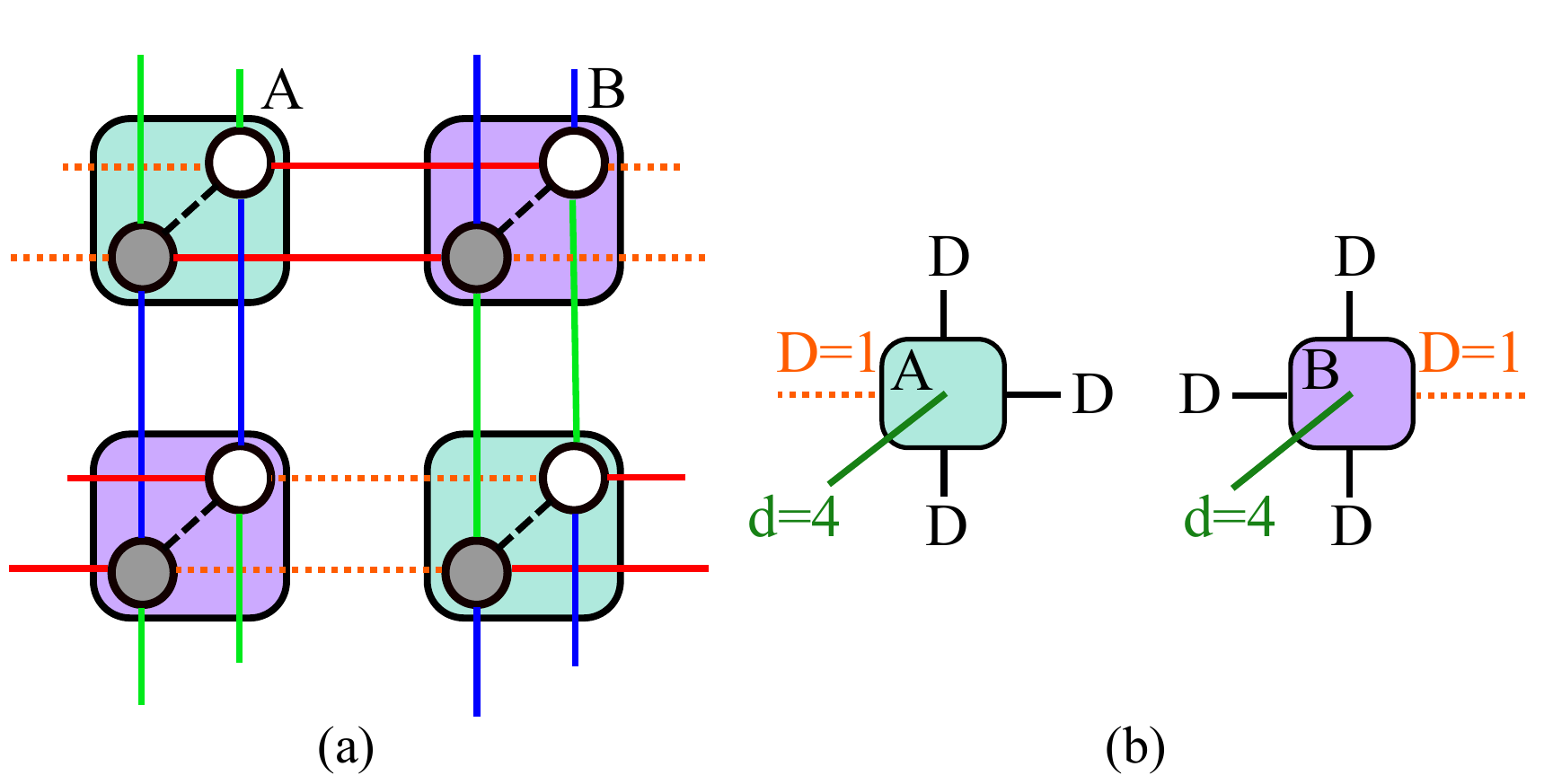}}
\caption{(Color online) (a) The $2\times 2$ iPEPS unit-cell with square geometry. Coarse-graining pairs of spins aligned vertically at site $i$ in upper and lower layers (shown by circles) creates distinct effective block sites represented by $A$ and $B$. Introducing trivial indices ($D=1$) denoted by orange dotted lines, converts the brick-wall honeycomb structure to a square lattice. (b) Associating iPEPS tensors to effective block sites. Solid dark green bonds correspond to physical indices with effective Hilbert space dimensions of $d=2^2$. }
\label{Fig:unitcell}
\end{figure}

In the extreme limit where $\theta=0\,(\pi)$, the model reduces to two decoupled honeycomb layers, each of which is governed by an AFM (FM) isotropic Kitaev exchange interaction and a gapless QSL ground state. In this regime, the model is exactly solvable in the Majorana fermion representation. Introducing a transformation from spin operators to Majorana fermion operators, $S_{i}^{\alpha}=i\gamma_{i,\alpha}^{x}\gamma_{i,\alpha}^{y}\gamma_{i,\alpha}^{z}$, one can map the Kitaev Hamiltonian to independent Majorana fermions and show that two different types of excitations arise: gapped magnetic vortices, as well as gapless Majorana fermions \cite{kitaev2006anyons}. 

\begin{figure*}
\centerline{\includegraphics[width=18cm]{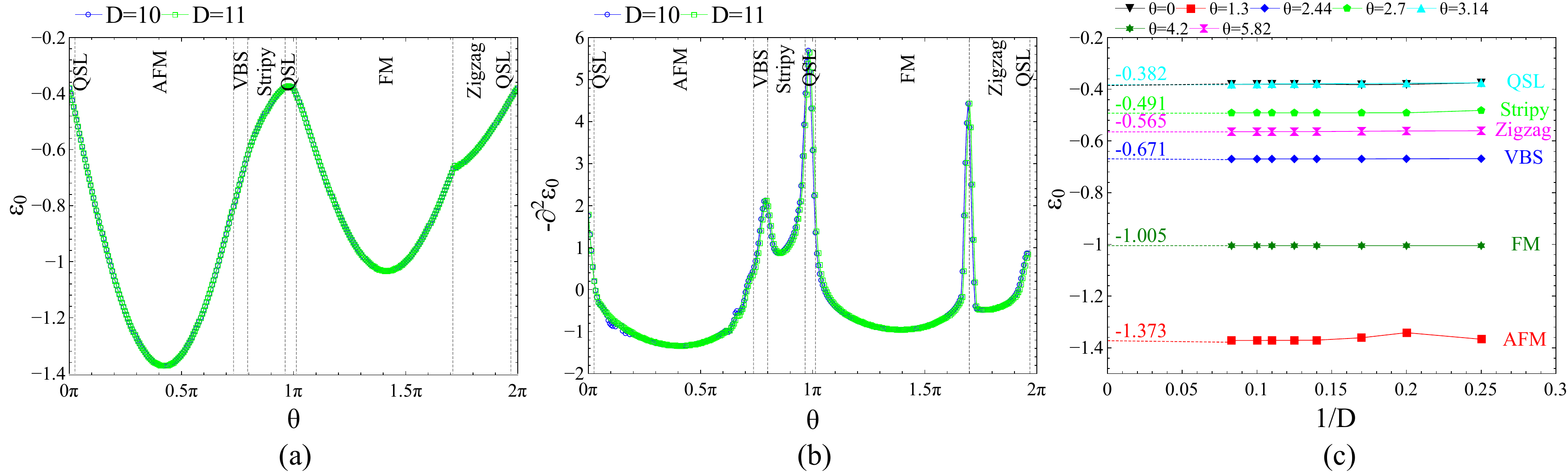}}
\caption{(Color online) (a) The ground-state energy per site $\varepsilon_{0}$ for the Kiatev-Heisenberg bilayer honeycomb lattice model versus coupling constant $\theta$ as obtained by iPEPS simulations for bond dimensions $D=10, 11$ and boundary dimension $\chi=64$ in the thermodynamic limit. (b) The second-order derivative of $\varepsilon_{0}$(c) Convergence of $\varepsilon_{0}$ within all phases with increasing bond dimension. The CTMRG boundary dimension is $\chi=D^2$. For $D\geq 8$, we set $\chi=64$. Values above dashed lines (linear fitting) mark the extrapolation of $\varepsilon_{0}$ as $D\rightarrow\infty$.}
\label{Fig:energy}
\end{figure*}

Additionally, the $\theta=-\frac{\pi}{2}$ regime corresponds to two decoupled FM Heisenberg Hamiltonians with ferromagnetic spin ordering. In this regime, the excitations of the model are commonly known as magnons (quantized collective spin waves) wherein low-energy states have a quadratic dispersion relation form as $\varepsilon_{k}\sim k^2$. On the other hand, the $\theta=\frac{\pi}{2}$ limit corresponds to the AFM Heisenberg Hamiltonian with an antiferromagnetic underlying ground state, giving rise to finite staggered magnetization. In this regime, the excitations are magnons with linear dispersion relation $\varepsilon_{k}\sim k$. 

Similar to the monolayer Kitaev-Heisenberg model, we also expect the bilayer to host mutual symmetric points that are related to each other by specific sub-lattice spin rotations that result in stripy and zigzag states from the FM and AFM states, respectively \cite{chaloupka2010kitaev,chaloupka2013zigzag}. In Sec.~\ref{sec:phase_diagram}, we will numerically confirm the existence of these phases in the phase diagram of the bilayer Kitaev-Heisenberg model in the thermodynamic limit.

\section{Methods}
\label{sec:method}
In the following we first describe the iPEPS technique and then give all relevant information to the applied high-order series expansions.

\subsection{The iPEPS Method}
The Projected Entangled Pair State (PEPS) technique \cite{verstraete2004renormalization} extends matrix product states \cite{Verstraete2008} to higher dimensions, offering a robust framework for simulating the ground states of quantum many-body systems (QMBS). By arranging tensors on a lattice where each represents a system site linked by entangled bonds, PEPS efficiently captures complex interactions and correlations, needed for characterizing the ground state wave function of a local Hamiltonian. The infinite version of PEPS (iPEPS) is further tailored to approximate the ground state of the QMBS on the translationally invariant infinite lattices in the thermodynamic limit \cite{Orus2009,Phien2015,jahromi_infinite_2018}. In this paper, we utilized the iPEPS approach based on the simple update (SU) algorithm \cite{Jiang2008} to simulate the spin-$1/2$ Kitaev-Heisenberg model on the infinite bilayer honeycomb lattice. Our simple update is armed with an imaginary-time evolution based on second-order Suzuki-Trotter decomposition of local Hamiltonian gates with the imaginary-time $\delta\tau$ that initiates from $10^{-1}$ and is gradually reduced to $10^{-3}$ while allowing a maximum of $3000$ iterations for each $\delta\tau$. Once the simple update is converged, the corner transfer matrix renormalization group (CTMRG) algorithm \cite{Nishino1996,Orus2009} is employed to contract the infinite tensor network and evaluate expectation values of local observables. Our maximum achievable bond dimension for iPEPS tensors of the bilayer KH model was $D=11$ which was achieved on a $4\times 2$ unit-cell. Although the smallest allowed TN unit-cell for the honeycomb lattice is $2\times 2$, we conducted our TN simulations on a larger unit-cell to ensure the detection of all potential phases characterized by non-local quantum correlation within the unit-cell. Additionally, we use boundary dimension $\chi=D^2$ for truncating the CTMRG environment tensors up to $D=8$. For $D>8$ we used $\chi=64$ due to limitation in computational resources. However, we observed that $\chi=64$ was already sufficient for the convergence of environmental tensors. 

The iPEPS technique was originally designed to simulate the ground state of the QMBS on the square lattices \cite{Phien2015}. However, one can apply the iPEPS method for representing quantum states on various two- or higher-dimensional lattice structures with proper choice of coarse-graining \cite{jahromi2018infinite} or fine-graining \cite{Schmoll2019FineGT}. In order to implement the iPEPS ansatz for the bilayer honeycomb lattice structure, a systematic approach involves mapping the nearest neighbor interactions on the bilayer honeycomb lattice to the nearest neighbor interactions on the square lattice. The process starts with grouping pairs of spins at sites $i$ of the upper and lower layers into effective block sites (see sites $A$ and $B$ in Fig.~\ref{Fig:unitcell}), and deforming the resulting coarse-grained lattice into a brick-wall structure. The resulting brick-wall lattice is topologically equivalent to a single-layer honeycomb lattice. The coarse-grained block sites have the local Hilbert space dimension of $d=2^2$, reflecting the total Hilbert space of two spins vertically aligned at site $i$ of the bilayer lattice. Next, by introducing trivial indices as orange dotted lines in Fig.~\ref{Fig:unitcell} and associating rank-$5$ tensors to each block site, an iPEPS ansatz for the square lattice with a $2\times 2$ unit-cell emerges that effectively represents the bilayer honeycomb lattice. Larger unit-cells can further be obtained by adding more tensors with the checker-board pattern as shown in Fig.~\ref{Fig:unitcell}.  Finally, the local two-body terms of the KH model on the effective square lattice are defined as explained in Appendix.~\ref{append:Hamiltonian}. These gates are later used in the simple update for simulating the ground state of the system in the thermodynamic limit.

\subsection{Series expansions} 
\label{sec:pCUT}

Complementary to the iPEPS technique, we perform high-order series expansions about the limit of isolated rung dimers where $J_{i,j}=K=0$ in order to explore the extent of the VBS phase. To this end we introduce the perturbation parameters $\lambda_J\equiv J_{i,j}/J_{iu,id}$ and $\lambda_K\equiv K/J_{iu,id}$ and set ${J_{iu,id}=1}$. 
The isotropic case of equally strong inter- and intralayer Heisenberg interactions is then recovered for $\lambda_J=1$ implying antiferromagnetic intralayer Heisenberg interactions $J_{i,j}>0$. 

The unperturbed limit $\lambda_J=\lambda_K=0$ corresponds to isolated Heisenberg dimers so that the product state of singlets on all rungs is the exact ground state and the excitations are spin-one triplets. 
At finite perturbation the elementary excitations of this VBS phase are then given by spin-one triplons (dressed triplets) \cite{Schmidt2003}.

We calculate the ground-state energy and the one-triplon dispersions as a perturbative series in $\lambda_{J,K}$ up to high orders.
Technically, the series expansion is realized with the help of perturbative continuous unitary transformations (pCUTs) \cite{Knetter2000,Knetter2003} which we describe in the following.

As the unperturbed part for $\lambda_J=\lambda_K=0$ is diagonal in the basis of singlets $\ket{s_i}$ and triplets $\ket{t_i^\alpha}$ with \mbox{$\alpha \in \{x,y,z\}$} on the rung dimers $i$, it can be written as
\begin{eqnarray}\label{h_0_q}
\mathcal{H}_0 &=& E_0+\mathcal{Q}\, ,
\end{eqnarray}
where $E_0=-(3/4)N_{\rm d}$ with $N_{\rm d}=N/2$ the number of rung dimers is the bare ground-state energy and \mbox{$\mathcal{Q}=\sum_{i,\alpha} t_i^{\alpha\dagger} t_i^{\alpha\phantom{\dagger}}$} with $\ket{t_i^\alpha}=t_i^{\alpha\dagger} \ket{s_i}$ is the counting operator of local triplet excitations. Nearest-neighbor rung dimers interact via the intralayer perturbations corresponding to $\lambda_J$ and $\lambda_K$. 
Using triplet operators, one can rewrite the bilayer Kitaev-Heisenberg model exactly as
\begin{equation}
\label{Eq:Hami_final}
{\cal H}={\cal H}_0+ \sum_{n=-2}^2 \hat{T}_n\, ,
\end{equation}
with $[\mathcal{Q},\hat{T}_n]=n\hat{T}_n$ and $n\in\{0,\pm 2\}$. Physically, the operator \mbox{$\hat{T}_n \equiv\sum_{j=J,K} \lambda_{j} \hat{T}^{(j)}_n$} corresponds to all processes where the change of energy quanta with respect to $\mathcal{H}_0$ is exactly $n$. 

In pCUTs, a Hamiltonian of the form \eqref{Eq:Hami_final} is mapped model-independently up to high orders in perturbation to an effective Hamiltonian $\mathcal{H}_\text{eff}$ with $[\mathcal{H}_{\rm eff},\mathcal{Q}]=0$. The general structure of $\mathcal{H}_{\rm eff}$ is then a weighted sum of operator products $\hat{T}_{n_1}^{(j_1)}\cdots \hat{T}_{n_k}^{(j_k)}$ in order $k$ perturbation theory. The block-diagonal $\mathcal{H}_\text{eff}$ conserves the number of triplon quasi-particles (qp). This represents a major simplification, since one can treat each triplon sector separately. The second part in pCUTs is model-dependent and corresponds to a normal-ordering of $\mathcal{H}_\text{eff}$. This is most efficiently done via a full graph decomposition in linked graphs using the linked-cluster theorem and an appropriate embedding scheme afterwards. Here, we focus on the zero- and one-triplon properties allowing to determine the ground-state energy and one-triplon excitation energies. 

The ground-state energy is obtained from $\langle 0
| \mathcal{H}_\text{eff} | 0 \rangle$, where $|0 \rangle = \prod_{i} | s_{i}
\rangle$ is the product state of isolated rung singlets. 
We have calculated series expansions of type $E_0/N_{\rm d} = \sum_{l+m\leq 8} a_{l,m} \lambda_J^{l} \lambda_K^{m}$ up to order 8.

For the one-triplon excitations, we employ translational invariance of the
honeycomb lattice with its underlying two-site basis. The dispersion
$\omega(\bm{k})_{\alpha,\mu}$
with $\alpha\in\{x,y,z\}$ referring to the three triplon flavors
and $\mu{=}1,2$ labeling the sublattices 
demands the diagonalization of 6$\times$6-matrices for each momentum $\bm{k}$. As the intralayer Kitaev- and Heisenberg interactions are flavor conserving these matrices are actually block-diagonal with three $2 \times 2$ blocks corresponding to the three triplon flavors.
 One then obtains six one-triplon bands which we determined up to order $k_\text{max}=8$
 in perturbation. 

The minimum of the six one-triplon bands is called the one-triplon gap $\Delta (\lambda)$ which is located at different momenta $\bm{k}_{\rm c}$ when tuning the ratio between $\lambda_J=\lambda \sin\theta$ and $\lambda_K=\lambda \cos\theta$. A gap-closing at the quantum-critical point $\lambda_{\rm c}$ corresponds to the breakdown of the rung-singlet VBS phase and signals the presence of long-range magnetic order for $\lambda >\lambda_{\rm c}$. The type of magnetic order is connected to the gap momentum. 

The location of the quantum-critical point $\lambda_{\rm c}$ can be extracted from the high-order series expansion by DlogPad\'{e} extrapolations. The Padé approximant $P[L, M]_{\Delta}$ of the quantity $\Delta(\lambda)$ is defined as
\begin{equation}
	P[L, M]_{\Delta} = \frac{P_L(\lambda)}{Q_{M}(\lambda)} = \frac{p_0 + p_1\lambda + \cdots + p_L\lambda^L }{1 + q_1\lambda + \cdots + q_M\lambda^M}\ ,
	\label{eq:pade}
\end{equation}
where $p_i, q_i \in \mathbb{R}$ and the degrees $L$, $M$ of the numerator polynomial $P_{L}(\lambda)$ and denominator polynomial $Q_{M}(\lambda)$ are restricted to $L+M\leq k_\text{max}$. 
Similarly, the Padé approximant of the logarithmic derivative of $\Delta$ is defined as
\begin{equation}
	P[L, M]_{\mathcal{D}} = \frac{{\rm d}}{{\rm d}\lambda} \ln(\Delta)
	\label{eq:dlogpade}
\end{equation}
with $L+M\leq k_\text{max} -1 $ and the associated DlogPadé approximant is obtained by integration. 

By analyzing the poles of the approximants and identifying the physical pole we can determine the quantum-critical point $\lambda_{\rm c}$.
In practice, we structure the DlogPadé extrapolants into families with $L-M=\text{const.}$ and calculate the average over the highest-order extrapolants of each family to determine $\lambda_{\rm c}$. We take the standard deviation of the average as a measure for the uncertainty of the extrapolation. 

\section{Phase Diagram} 
\label{sec:phase_diagram}
In this section, we study the full phase diagram of the bilayer Kitaev-Heisenberg model, i.e, $H=H_{KH}+H_{H}$. To this end, we set the Kitaev and Heisenberg couplings equal to $K=\cos\theta$ and $J=\sin\theta$, respectively and approximate the ground state of the KH model with the iPEPS technique over the full interval $\theta\in[0,2\pi]$. Once the ground state is obtained for different bond dimensions $D$, an accurate analysis of the phase diagram becomes possible through the calculation of energy and expectation values of relevant physical observables.
 \begin{figure}
\centerline{\includegraphics[width=\columnwidth]{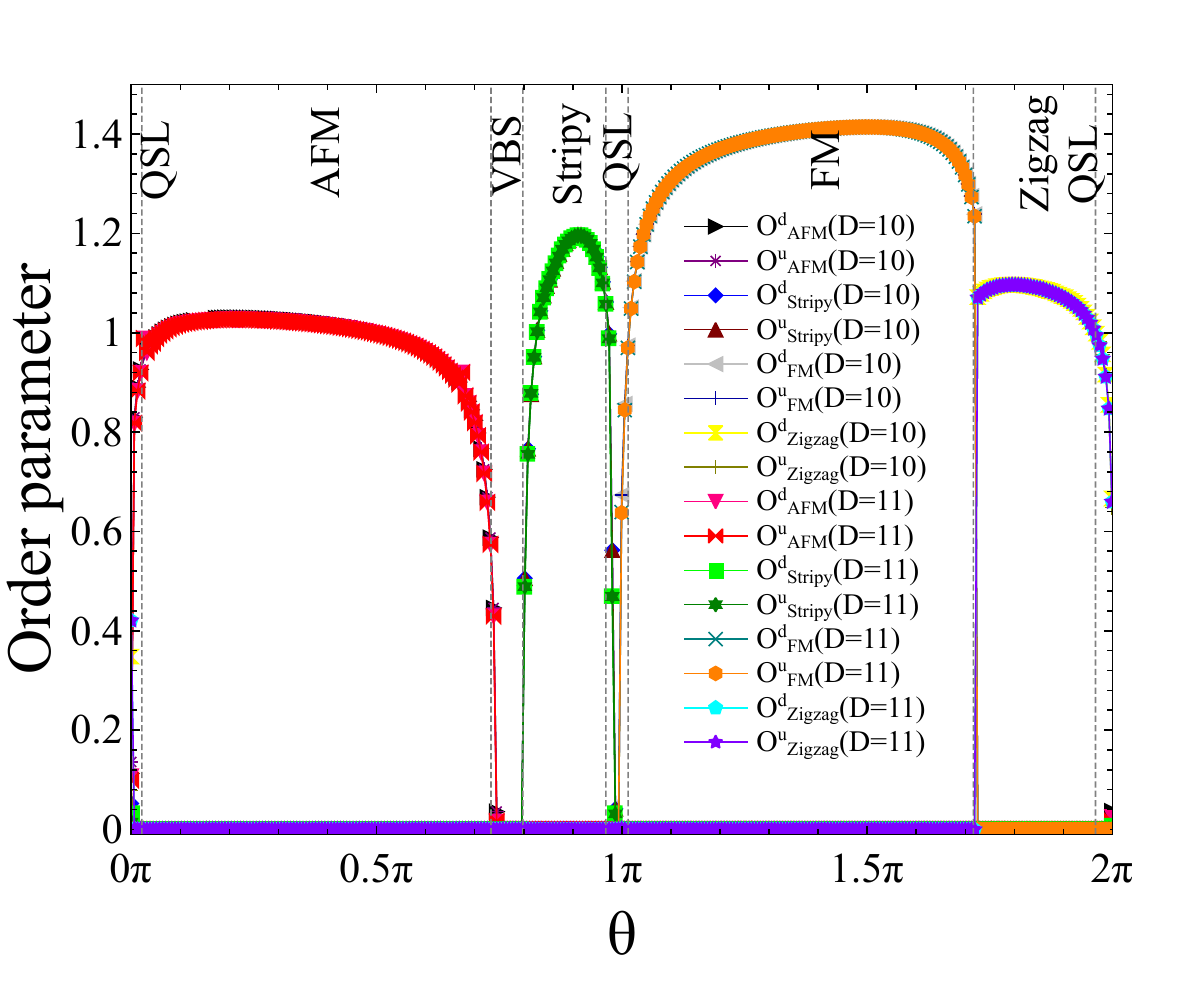}}
\caption{(Color online) Different order parameters for the magnetically ordered phases obtained by iPEPS simulations on the bilayer honeycomb lattice with bond dimensions $D=10,11$ and boundary dimension $\chi=64$. $O^{d}_{l}, O^{u}_{l}$ refer to the order parameter of lower and upper layers of phase $l$, respectively.}
\label{Fig:order}
\end{figure}

We first calculated the ground-state energy $\varepsilon_0$ and its second-order derivative to capture different phases and pinpoint their transition points. We unveil the presence of seven distinct phases in the thermodynamic limit. Fig.~\ref{Fig:energy}-(a) and (b) show the ground-state energy per site, $\varepsilon_0$, as well as its second-order derivative obtained with iPEPS simulations for bond dimensions $D=10, 11$. The phase boundaries are detected more precisely in the derivative plot. Fig.~\ref{Fig:energy}-(c) further demonstrates the scaling of ground-state energy versus inverse bond dimension $D$ for representative points deep inside each phase. The excellent convergence within all phases suggests that the underlying phases are stable in the thermodynamic limit and are not artifacts of finite bond dimension. 

To gain a proper understanding of different phases, it is useful to delve into extreme $\theta$ values, where the model simplifies to the pure Kitaev (Heisenberg) Hamiltonian. As previously pointed out, at $\theta=0$ ($\theta=\pi$) the KH model is reduced to two decoupled AFM (FM) Kitaev Hamiltonians with gapless Kitaev QSL ground states within each layer. Notably, the zero total magnetization in the TN results is in alignment with the nature of QSL state at these extreme $\theta$ values. However, our further analysis of energy reveals that the extended AFM (FM) QSL spans in the range $\theta\in[359^\circ,2^\circ] 
  ([175^\circ,182^\circ])$, see Fig.~\ref{Fig:energy}-(a), ~\ref{Fig:energy}-(b). 
 
 In the opposite extreme limit, where $\theta=-\frac{\pi}{2}$, the FM Heisenberg Hamiltonian governs the system, inducing long-range FM order characterized by finite magnetization. To effectively detect the FM order, we used an order parameter defined as $(\sum_{i}\langle \mathbf{S}_i \rangle)^2$. Our simulations unveil the appearance of FM order within both layers in the angular interval $\theta\in[182^\circ,311^\circ]$, including $\theta=-\frac{\pi}{2}$. Our analysis of local magnetization further reveals the parallel alignment of upper and lower layers. At $\theta=\frac{\pi}{2}$ we are left with the AFM Heisenberg Hamiltonian on the bilayer lattice.  The ground state of the system in this limit has AFM ordering with finite staggered magnetization. This ordering is quantified by the order parameter, defined as  $(\sum_{i}(-1)^{i}\langle \mathbf{S}_i \rangle)^2$. Our TN results further show that the AFM ordered ground state persists in both layers across a broad range of $\theta$, spanning $[2^\circ,135^\circ]$ which also includes $\theta=\frac{\pi}{2}$. Besides, we find that upper and lower layers have the anti-parallel alignment pattern of magnetization within this phase.

\begin{figure*}
\centerline{\includegraphics[width=18cm]{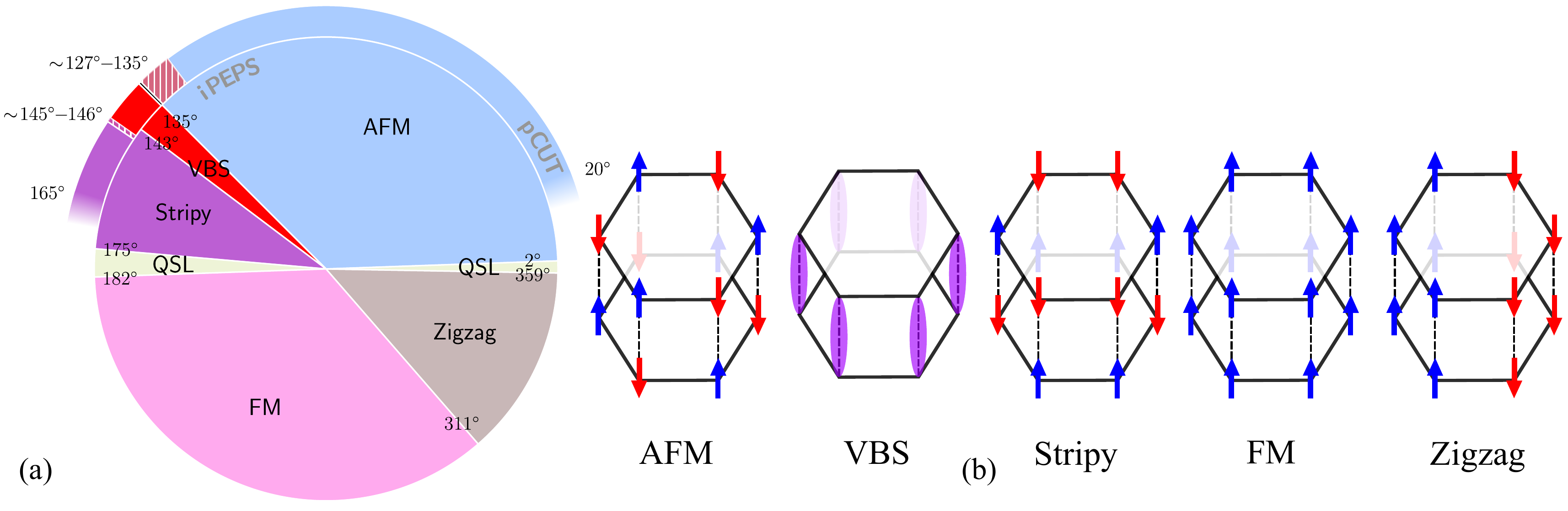}}
\caption{(Color online) (a) The ground state phase diagram of the spin-$1/2$ Kitaev-Heisenberg model on the bilayer honeycomb lattice obtained from iPEPS and pCUT. Striped regions indicate the uncertainty of pCUT in the determination of the phase boundaries. The black solid line refers to $\theta=135^\circ$ signaling the change of gap momentum between AFM and stripy orders. (b) Magnetization patterns of various ordered phases for both layers. Upper (lower) honeycombs correspond to upper (lower) layers. Within the AFM and stripy ordered phases, upper and lower layers exhibit antiparallel alignment, characterized by opposing magnetic orientation, whereas at FM and zigzag phases neighboring layers have parallel magnetic orientation. The rung singlets in the VBS phase have been further illustrated by thick purple links connecting the two layers.}
\label{Fig:phase}
\end{figure*}

Beyond previously discussed phases, including QSL states as well as FM and AFM magnetically ordered states our simulations suggests the existence of three additional phases in the phase diagram. Earlier studies provided a mapping and showed that in addition to four aforementioned phases, the ground state of the monolayer Kitaev-Heisenberg model manifests two distinctive arrangements of spins associated with zigzag and stripy patterns \cite{chaloupka2010kitaev,chaloupka2013zigzag}. Our TN analysis of local magnetization shows that the zigzag ordering exist in the bilayer model as well, and broadly extends over  $\theta\in[311^\circ,359^\circ]$ with parallel alignment of upper and lower layers. In order to capture this ordering, we define the zigzag order parameter as $(\sum_{i}\langle \mathbf{S}_i \rangle -\langle\mathbf{S}_{i+1} \rangle)^2$, where $\mathbf{S}_i$ and $\mathbf{S}_{i+1}$ represent two neighboring spins along the $x$-direction on the $4\times2$ iPEPS unit-cell. On the other hand, when $\theta$ spans $[143^\circ,175^\circ]$, our tensor network results show that the ground state hosts a stripy magnetic state with the anti-parallel alignment of upper and lower layers, which is best detected with stripy order parameter defined as $(\sum_{i}(-1)^{i}(\langle \mathbf{S}_i \rangle +\langle\mathbf{S}_{i+1} \rangle))^2$. Here again $\mathbf{S}_i$ and $\mathbf{S}_{i+1}$ are two neighboring spins along the $x$-direction.  Fig.~\ref{Fig:order} demonstrates the order parameters for upper and lower layers obtained by iPEPS simulations for maximum bond dimensions $D=10,11$. Moreover, the full phase diagram of the system as well as the sketches of the local ordering in the magnetic phases have been depicted in Fig.~\ref{Fig:phase}-(a) and (b), respectively. 

Thus far, we have identified six phases of the bilayer Kitaev-Heisenberg model. It is crucial to remind that aforementioned phases cover the full phase diagram of the monolayer Kitaev-Heisenberg model. In contrast, there is a relatively narrow phase within the interval $\theta\in[135^\circ,143^\circ]$,  which is absent in the monolayer counterpart. Our results show the vanishing local magnetization within this interval, suggesting the absence of conventional long-range ordering. Furthermore, analysis of the nearest neighbor spin-spin correlations on each of the honeycomb layers uncovers a uniformity across all bonds, as evidenced by the finite and equal value of $\langle\mathbf{S}_{i} \mathbf{S}_{j} \rangle$, which highlights the absence of any symmetry breaking. However the strong intralayer spin-spin correlation between the two layers suggests the existence of a phase with isolated dimers which is the characteristic of a VBS state.

In order to identify this new phase as a VBS, we supplement the iPEPS calculations with high-order series expansions as introduced in Sec.~\ref{sec:pCUT}. We restrict the discussion to the relevant parameter range $\theta \in  (0^\circ,180^\circ)$. Specifically, as a function of $\theta$, we determine the critical gap momentum $\bm{k}_{\rm c}$  and we analyze the closing of the one-triplon excitation gap $\Delta$. 

There are two distinct regimes regarding the critical gap momentum $\bm{k}_{\rm c}$ as a function of $\theta$. For $\theta \in (0^\circ,135^\circ]$, one has $\bm{k}_{\rm c}=(0,0)$ and antisymmetric eigenvectors $(1,-1)$ in the two-site unit cell, corresponding to antiferromagnetic Néel order (compare AFM in Fig.~\ref{Fig:phase}). We note that the gap is three-fold degenerate regarding the three triplon flavors.  
For $\theta \in [136^\circ,180^\circ)$, we find $\bm{k}_{\rm c}=(\pi,0)$ for the $t^x$ triplon excitation, $\bm{k}_{\rm c}=(0,\pi)$ for $t^y$, and $\bm{k}_{\rm c}=(\pi,\pi)$ for $t^z$. The eigenvectors take the form $(1,-1)$ for $t^x$ and $t^y$ and $(1,1)$ for $t^z$. The resulting ordering patterns correspond to stripy orders along the $x$, $y$ and $z$ interaction axes of the honeycomb lattice, respectively (see also the illustrated stripy order in Fig.~\ref{Fig:phase}-(b) exemplarily for stripy order along the $z$-interaction axis.). As expected for isotropic Kitaev interactions $K^\alpha = K$, all three excitations have an identical dispersion up to rotations of $120^\circ$ and it is sufficient to consider only one flavor. 	
We stress that the change of gap momentum from $\theta=135^\circ$ to $136^\circ$ perfectly coincides with the change from AFM order to the VBS phase at $\theta=135^\circ$ from iPEPS. 

Next we investigate the closing of the one-triplon excitation gap $\Delta(\lambda)$ at a critical $\lambda_{\rm c}$ to explore the stability of the VBS and its quantum-critical breakdown towards magnetically ordered phases. For the isotropic case $J_{i,j}=J_{iu,id}$, for which the phase diagram is depicted in Fig.~\ref{Fig:phase}-(a), one has $\lambda_J=1$ and therefore $\lambda=1/\sin \theta\equiv \lambda_{\rm iso}$. A critical point $\lambda_{\rm c} < \lambda_{\rm iso}$ implies a phase transition for $\lambda_J <1$ to a magnetically ordered phase which we expect to be stable up to $\lambda_J =1$. In contrast, for $\lambda_{\rm c} > \lambda_{\rm iso}$, one has a VBS adiabatically connected to the rung-singlet product state for the isotropic case. Here we assume that there is no first-order phase transition which we checked explicitly by comparing the ground-state energies of iPEPS and pCUT as a function of $\lambda$ (see Appendix \ref{append:gsenergy}).

	For the regime with $\bm{k}_{\rm c}=(0,0)$, we find $\lambda_{\rm c} < \lambda_{\rm iso}$ for $\theta\leq 127^\circ - 135^\circ$. This implies that for $\lambda_{\rm iso}$ depicted in Fig.~\ref{Fig:phase}-(a) we find an antiferromagnetically ordered phase, coinciding with the AFM phase found by iPEPS.
	The uncertainty in the critical $\theta$ originates from the uncertainty of the series extrapolation to locate the gap closing at $\lambda_{\rm c}$ (see also Sec.~\ref{sec:pCUT}).  
	
	For the other regime with $\bm{k}_{\rm c}=(\pi,\pi)$ for $t^z$, we find $\lambda_{\rm c} < \lambda_{\rm iso}$ for $\theta \in [146^\circ , 165^\circ]$. For $\theta \gtrsim 165^\circ$, the extrapolation of the one-triplon gap does not indicate a closing of the gap. However, we stress that $\lambda_{\rm iso}$ shifts to larger values when increasing $\theta$ towards $180^\circ$ so that the extrapolation of the gap series becomes unreliable when approaching the QSL phase detected by iPEPS.  
	Finally, for $\theta \in [136^\circ , 145^\circ]$, we find a critical momentum corresponding to stripy magnetic order but no closing of the one-triplon gap before $\lambda_{\rm iso}$, establishing the presence of a VBS phase which is adiabatically connected to the limit of isolated rung singlets. 
	
To sum up, we have mapped out the full phase diagram of the bilayer KH model in the thermodynamic limit and identified seven distinct phases. We have shown that similar to the monolayer model four magnetic ordered phases and two extended QSL states emerge as a result of the competition between the Heisenberg and Kitaev interactions. However, a distinctive feature of the bilayer model unfolds as the appearance of a VBS state within a narrow range of the parameter space. This phase which is positioned between the stripy and AFM magnetically ordered phases stands in sharp contrast to the monolayer model.

\section{Conclusions and discussion} 
\label{sec:conclude}
Real Kitaev materials frequently manifest deviations from the pure Kitaev model. The Kitaev-Heisenberg model emerges as an anisotropic spin-orbital model which can describe the low-energy physics of quasi-2D compounds such as Na$_{2}$IrO$_{3}$ and Li$_{2}$IrO$_{3}$. However, the layered structure of such materials may affect the overall magnetic behavior. In this regard, we studied the full phase diagram of spin-$1/2$ Kitaev-Heisenberg model on an infinite bilayer honeycomb lattice, where layers are coupled via Heisenberg interactions. 

We used the infinite projected entangled pair state tensor network algorithm to simulate the ground state of the model in the thermodynamic limit. In addition, we set up series expansions about the limit of isolated rung dimers and determined the ground-state energy as well as one-triplon excitation energies up to high order in perturbation. The results derived from these two complementary techniques demonstrate that seven distinct phases cover the full phase diagram, including four magnetically ordered phases as well as two extended QSL states and a VBS state. While the quantum phase diagram of the model bear a remarkable resemblance to the monolayer counterpart, layering the model introduces a stark contrast, as evidenced by the emergence of the VBS state in a bilayer lattice, a relatively narrow phase within the parameter space which one does not observe in the monolayer system.
Our study shows that understanding the impact of layered structures on the emergence of novel quantum phases is crucial for advancing the field of condensed matter physics. Besides, it may open new avenues for designing materials with specific quantum properties for future technologies. 
  
\section{Acknowledgements}
S.S.J. acknowledges the support from the Institute for Advanced Studies in Basic Sciences (IASBS). K.P.S. gratefully acknowledge financial support by the Deutsche Forschungsgemeinschaft (DFG, German Research Foundation) through the TRR 306 QuCoLiMa ("Quantum Cooperativity of Light and Matter") - Project-ID 429529648 (KPS). KPS acknowledges further financial support by the German Science Foundation (DFG) through the Munich Quantum Valley, which is supported by the Bavarian state government with funds from the Hightech Agenda Bayern Plus. P.A., A.D., and K.P.S. gratefully acknowledge the scientific support and HPC resources provided by the Erlangen National High Performance Computing Center (NHR@FAU) of the Friedrich-Alexander-Universität Erlangen-Nürnberg (FAU) under the NHR project b177dc (SELRIQS). NHR funding is provided by federal and Bavarian state authorities. NHR@FAU hardware is partially funded by the German Research Foundation (DFG) – 440719683.

\appendix
\section{iPEPS implementation of the Kitaev-Heisenberg model on the bilayer honeycomb lattice}
\label{append:Hamiltonian}
In Sec.~\ref{sec:method}, we explained how one can develop the square iPEPS ansatz for bilayer honeycomb lattice structure. The basic idea for simulating honeycomb lattice with iPEPS technique has already been worked out in Ref.~\cite{jahromi2018infinite}. In what follows, we extend the idea to bilayer models. As pointed out previously, distinct block sites in a checkerboard pattern are created by coarse-graining pairs of spins from the upper and lower layers of the bilayer honeycomb lattice.  
 \begin{figure}
\centerline{\includegraphics[width=\columnwidth]{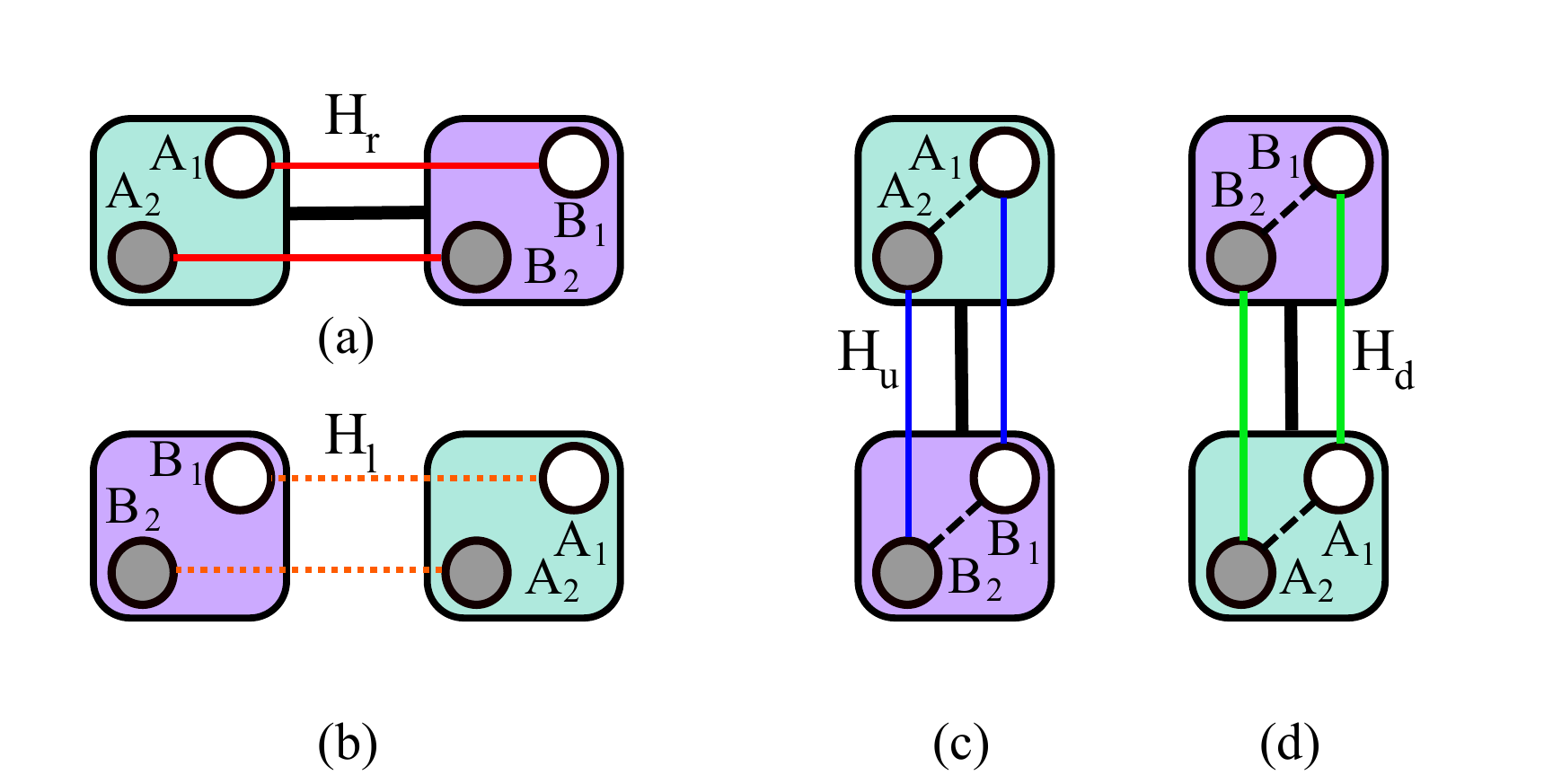}}
\caption{(Color online) Nearest-neighbor block sites, $A$ and $B$ with a checkerboard pattern in a $2 \times 2$ unit-cell. The spins within block sites are labeled by $A_1, A_2, B_1, B_2$ for applying the Hamiltonian of the bilayer model in the iPEPS formalism. $H_{r}, H_{l}, H_{u}$ and $H_{d}$ account for two-body local terms of Hamiltonian. (a) $H_r$ corresponds to the Hamiltonian along the $x$-direction connecting block site $A$ to $B$. (b) $H_l$ is the identity operator which acts trivially on horizontal links connecting block site $B$ to $A$, accounting for no interaction in this direction. (c) and (d) demonstrate the $H_u$ and $H_d$ interactions that acts on the vertical bonds of block sites along the $A$ to $B$ and alternatively, $B$ to $A$ directions.}
\label{Fig:hamilton}
\end{figure}
To assign a Hamiltonian to the block sites, we first label the two spins within the block sites $A (B)$ as $A_{1}, A_{2} (B_{1}, B_{2})$, see Fig.~\ref{Fig:hamilton}. Hence, the Hilbert space of two neighboring block sites (which is equivalent to a bond on the coarse-grained lattice) is given by
\be
\label{eq:A1}
\mathcal{H}_{\text{cell}} = \mathcal{H}_{A} \otimes \mathcal{H}_{B},
\ee
with local Hilbert space of each block site defined as:
\begin{align}
&\mathcal{H}_{A} = \mathcal{H}_{A_{1}} \otimes \mathcal{H}_{A_{2}} \label{eq:A2} \\
&\mathcal{H}_{B} = \mathcal{H}_{B_{1}} \otimes \mathcal{H}_{B_{2}}, \label{eq:A3}
\end{align}
where the local physical dimension of $\mathcal{H}_{A}$ and $\mathcal{H}_{B}$ is \mbox{$2^{2}=4$}. The Hamiltonian of the bilayer honeycomb model on the coarse-grained square lattice of block sites, therefore reads: 
\be
\label{eq:hblock}
\mathcal{H}_{\text{Bilayer}}=\sum_{i}\mathcal{H}_{i}+\sum_{\langle i j \rangle}\mathcal{H}_{i,j}^{x}+\sum_{\langle i j \rangle}\mathcal{H}_{i,j}^{y}
\ee
where $\langle i j \rangle$ refers to nearest neighbor block sites and $\mathcal{H}_{i}$ corresponds to local terms in the Hamiltonian. $\mathcal{H}_{i,j}^{x} (\mathcal{H}_{i,j}^{y})$ denotes the part of the Hamiltonian that acts on horizontal (vertical) bonds of the coarse-grained lattice. The local and nearest-neighbor interactions  terms are further defined as follows:
\begin{align}
&\mathcal{H}_{i} = h_{iA} + h_{iB},\\
&\mathcal{H}_{i,j}^{x} = h_{iA,jB}^{x} + h_{iB,jA}^{x},\\
&\mathcal{H}_{i,j}^{y} = h_{iA,jB}^{y} + h_{iB,jA}^{y},
\end{align}
where 
\begin{align}
&h_{iA} = J_{iu,id} (\mathbf{S}_{A_{1}} \cdot \mathbf{S}_{A_{2}}),\\
&h_{iB} = J_{iu,id} (\mathbf{S}_{B_{1}} \cdot \mathbf{S}_{B_{2}}),
\end{align}
and nearest neighbor terms along the $x$ direction correspond to
\begin{align}
h_{iA,jB}^{x} &= J_{i,j}(\mathbf{S}_{A_{1}} \cdot \mathbf{S}_{B_{1}}) + K^{z}(S_{A_{1}}^{z}S_{B_{1}}^{z}) \nonumber \\
              &\quad + J_{i,j}(\mathbf{S}_{A_{2}} \cdot \mathbf{S}_{B_{2}}) + K^{z}(S_{A_{2}}^{z}S_{B_{2}}^{z}), \\
h_{iB,jA}^{x} &= \mathbb{I}.
\end{align}

This is while one can write the two-body terms along the $y$ direction as 
\begin{align}
h_{iA,jB}^{y} &= J_{i,j}(\mathbf{S}_{A_{1}} \cdot \mathbf{S}_{B_{1}})+K^{x}(S_{A_{1}}^{x}S_{B_{1}}^{x})\nonumber \\
              &\quad + J_{i,j}(\mathbf{S}_{A_{2}} \cdot \mathbf{S}_{B_{2}}) + K^{x}(S_{A_{2}}^{x}S_{B_{2}}^{x}), \\
h_{iB,jA}^{y} &= J_{i,j}(\mathbf{S}_{A_{1}} \cdot \mathbf{S}_{B_{1}})+K^{y}(S_{A_{1}}^{y}S_{B_{1}}^{y}) \nonumber \\
              &\quad + J_{i,j}(\mathbf{S}_{A_{2}} \cdot \mathbf{S}_{B_{2}}) + K^{y}(S_{A_{2}}^{y}S_{B_{2}}^{y}).
\end{align}
Finally, we can write all two-body terms applied in imaginary time evolution in the iPEPS formalism as
\begin{align}
&\mathcal{H}_{r} = h_{iA,jB}^{x},\\
&\mathcal{H}_{l}=\mathbb{I},\\
&\mathcal{H}_{u}=\frac{1}{2}(h_{iA}+h_{iB})+h_{iA,jB}^{y},\\
&\mathcal{H}_{d}=\frac{1}{2}(h_{iA}+h_{iB})+h_{iB,jA}^{y},
\end{align}
where $\mathcal{H}_{r}, \mathcal{H}_{l}, \mathcal{H}_{u}, \mathcal{H}_{d}$ are shown in the Fig.~\ref{Fig:hamilton}.

\section{Comparison of ground-state energies from iPEPS and pCUT in the VBS phase}
\label{append:gsenergy}
\begin{figure}
	\includegraphics[width=\columnwidth]{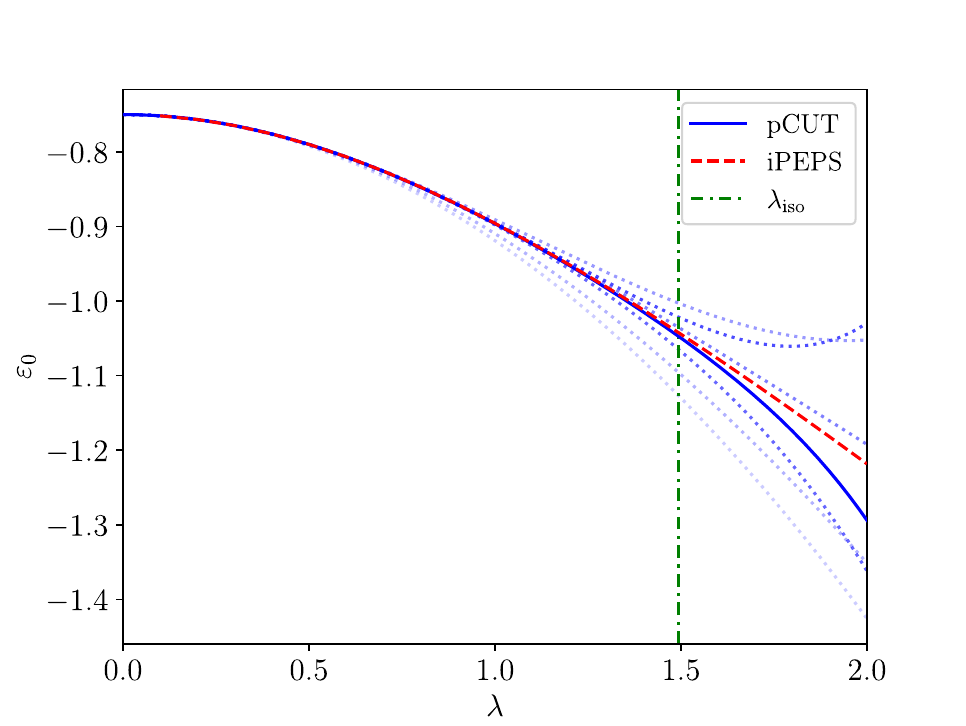}
	\caption{Ground-state energy per rung dimer in the VBS phase (exemplarily for $\theta=138^\circ$) from iPEPS and pCUT calculations as a function of $\lambda$. The iPEPS results are shown as red dashed line. The highest-order pCUT results are shown as blue solid line with blue dotted lines showing the results from lower orders with decreasing opacity. The green dashed-dotted line indicates the value $\lambda_{\rm iso}$ of isotropic inter- and intralayer Heisenberg interactions. }
	\label{fig:gsenergy}
\end{figure}

In this section we compare and analyze the ground-state energies obtained from iPEPS and pCUT calculations. 
The ground-state energy per rung dimer, $\varepsilon_0$, is plotted as a function of $\lambda$ in Fig.~\ref{fig:gsenergy} exemplarily for $\theta=138^\circ$. For this value of $\theta$, both pCUT and iPEPS clearly indicate the presence of a VBS phase (compare Fig.~\ref{Fig:phase}). To recall the definition of $\lambda$, the ratios between the intra- and interlayer interaction strengths are given by ${K/J_{iu,id}=\lambda \cos\theta}$ and $J_{i,j}/J_{iu,id} = \lambda \sin\theta$. Hence, the isotropic case $J_{i,j} = J_{iu,id}$ described in Sec.~\ref{sec:phase_diagram} is obtained for ${\lambda = \lambda_{\rm iso} = 1/\sin\theta}$.

The iPEPS simulations were performed with bond dimension $D=6$. The pCUT results were calculated up to order 8 in $\lambda$ and are shown as bare (i.\,e. not extrapolated) series. Additionally, lower orders are plotted to illustrate the convergence behavior of the series. Up to $\lambda\approx 0.8$ both methods agree almost perfectly with each other. Although the deviations between the methods increase with $\lambda$ as expected, even up to $\lambda_{\rm iso}$ the qualitative behavior stays the same. The obtained results do not indicate a first-order phase transition up to the isotropic point $\lambda_{\rm iso}$. Thus we expect the rung-singlet phase found for $\lambda=0$ to be stable until the point of a second-order phase transition as discussed in Sec.~\ref{sec:phase_diagram}.

\bibliography{references}{}
\bibliographystyle{apsrev4-1}

\end{document}